**Author ID Page**

Cassandra Crompton, PharmD- *corresponding author*

Penn Presbyterian Medical Center Department of Pharmacy

Philadelphia, PA

Kassidy Baum, PharmD

Penn Presbyterian Medical Center Department of Pharmacy

Philadelphia, PA

Michael Silvey, PharmD

Penn Presbyterian Medical Center Department of Pharmacy

Philadelphia, PA

Mona Al Mukaddam, MD, MS

Penn Presbyterian Medical Center

Philadelphia, PA

**There are no conflicts of interest to disclose.**

# Comparison of Forteo® and Teriparatide on Improvements in Bone Mineral Density

**Abstract**

*Purpose:*

PF708 teriparatide, manufactured by Alvogen, was approved through the 505(B)(2) application after the expiration of Forteo® market exclusivity in October 2019. The 505(B)(2) application allows safety and efficacy information to come from studies not conducted by the applicant if it is demonstrated that the proposed product shares characteristics (active ingredient, dose, route of administration, strength) with the reference product and comparable pharmacokinetic data is established. Real-world data is thus needed to demonstrate equivalent safety and efficacy between Forteo® and PF708 teriparatide. The purpose of this study was to evaluate changes in bone mineral density (BMD) in osteoporotic patients treated with either Forteo® or PF708 teriparatide for 18-24 months.

*Methods:*

This retrospective chart review included osteoporotic patients ≥ 18 years of age who received PF708 teriparatide or Forteo® prescriptions from the practice location and initiated therapy between October 1, 2019 and October 31, 2022. Patients without DXA results at the conclusion of therapy or who did not complete at least 18 months of therapy were excluded. The primary endpoint was percent change from baseline in BMD of the lumbar spine, femoral neck, and total hip. Data was analyzed using a two-tailed Mann Whitney test with a significance level of 0.05.

*Results:*

A total of 108 patients were included: 27 (median age 67 years, 88.9% female) in the PF708 teriparatide group and 81 (median age 68 years, 85.2% female) in the Forteo® group. There was no significant difference in median change from baseline BMD of lumbar spine (+9.6% vs +7.7%, P= 0.56), left femoral neck (1.5% vs 2.8%, P= 0.39), right femoral neck (+1.6% vs +4.4%, P= 0.14), left total hip (+2.4% vs +3.2%, P= 0.24), or right total hip (0.0% vs +1.4% , P= 0.71) with PF708 teriparatide vs Forteo®, respectively. Adverse effects were reported in 7.4% of patients in teriparatide group and 9.9% of patients in the Forteo® group. Most common reported adverse effects were constipation, nausea, and fatigue.

*Conclusion:*

In this small retrospective study, effect on BMD and reported adverse events were similar between PF708 teriparatide and Forteo® when used for the recommended duration of therapy. These results contribute real-world data to support equivalent efficacy and safety of PF708 teriparatide and Forteo® formulations.

**Background**

Osteoporosis is an age-related condition characterized by decreased bone strength and quality leading to an increased risk of fractures.[1] Osteoporosis is more prevalent in women than men in the United States, which is attributed to the loss of ovarian function at menopause. [1] Osteoporosis-related fragility fractures are defined as fractures that occur from low energy trauma such as a fall from standing height. [1] Fragility fractures of the spine and hip are associated with increased risk of mortality; one-year mortality rate in older adults with osteoporotic fractures is approximately 15% for women and 22% for men.[2] Fracture risk increases with previous history of fracture, history of fracture in first-degree relatives, current cigarette smoking, malnutrition, alcoholism, inadequate physical activity, and poor overall health.[1]

Bone mineral density (BMD) is determined by the balance of bone resorption and formation; a process known as bone remodeling. Bone resorption is the process in which osteoclasts break down bone tissue resulting in the release of calcium in the blood. Bone formation is the synthesis of new bone tissue by osteoblasts. The rate of bone remodeling is regulated by estrogens, androgens, vitamin D, and parathyroid hormone (PTH) and serves to repair microdamage within the skeleton and maintain serum calcium levels. Osteoporosis develops when bone resorption exceeds formation resulting in net loss of bone. [1,3] Bone mineral density is monitored with the use of dual energy X-ray absorptiometry (DXA)[4]

Teriparatide is a recombinant formulation of endogenous PTH that is used to increase osteoblast function, gastrointestinal calcium absorption, and renal calcium absorption to increase BMD, bone mass, and strength.[5] The European Society of Endocrinology recommends the use of teriparatide, and other PTH analogs, for postmenopausal women with osteoporosis at high or very high risk of fracture.[6]

Teriparatide (Forteo®) was FDA approved in November 2002 for the treatment of men and postmenopausal women with osteoporosis and high risk for fracture and in July 2009 for treatment of glucocorticoid-induced osteoporosis.[4] Market exclusivity of teriparatide (Forteo®) expired in October 2019 allowing for biosimilars to be introduced.[7] A biosimilar is a biologic medication (e.g., vaccine, blood and blood components, gene therapy, recombinant therapeutic proteins) that is highly similar and has no clinically meaningful difference from the reference product.[8] Approval is obtained through completion of an abbreviated biologic license application (BLA).[8] The biosimilar approval pathway differs from approval of generic medications in that generic manufacturers need only demonstrate that the generic is bioequivalent to the brand name medication.[8] This process is achieved through completion of a 505(B)(2) application which allows safety and efficacy information required for approval to come from studies not conducted by or for the applicant.[8] Teriparatide (Forteo®) is defined as a recombinant protein therefore classifying it as a biologic.[5]

United States spending on prescription drugs has increased throughout the years due to increasing popularity of biologic therapies. It is anticipated that with the introduction of biosimilars, prescription drug spending on biologics will decrease by approximately $54 billion by 2026. These anticipated cost savings motivates insurance companies to update their formularies with preference for alternative formulations. This impacts patients who are stable on biologic therapy and need to switch to biosimilars due to changes in insurance coverage.

Switching therapy to insurance preferred biosimilars requires reliance on available safety and efficacy data to support clinical decision-making.[9]

Since the expiration of Forteo® market exclusivity, several teriparatide biosimilars have been introduced through the 505(B)(2) application process. Retrospective and prospective studies have been conducted in Asia and Europe to compare safety and efficacy between Forteo® and biosimilars and have shown no differences with respect to pharmacokinetics, pharmacodynamics, T-scores, and fracture risk.[10-14] A Center for Drug Evaluation and Research application for teriparatide was approved on October 4, 2019. The approved teriparatide product, considered a “follow-on” teriparatide, is identified as PF708 and manufactured by Alvogen. For purposes of clarity for this publication, PF708 will be identified as “PF708 teriparatide”. The application for approval included a randomized study comparing the effects of PF708 teriparatide and Forteo® in patients with osteoporosis. The primary objective was to compare the effects of PF708 teriparatide and Forteo® on immunogenicity. The secondary objective was to compare the pharmacokinetic and pharmacodynamic effects of PF708 teriparatide and Forteo®, which endpoints included mean change from baseline in lumbar spine BMD after 24 weeks of treatment. Results of the primary objective found no clinically significant difference in percentage of subjects with anti-drug antibodies between PF708 teriparatide and Forteo®. Results of the secondary objective were not published due to the following: “the design of the trial was not adequate to allow comparison of BMD and bone turnover marker results between PF708 and Forteo®”. Therefore, due to the 505(B)(2) approval pathway, to our knowledge there is currently no available real-world outcomes data comparing Forteo® and the PF708 teriparatide formulation.[7]

## Methods

This retrospective chart review included patients ≥18 years of age diagnosed with osteoporosis who received prescriptions from the practice location and were initiated on Forteo® or PF708 teriparatide between October 1, 2019 and October 31, 2022. Patients without DXA results at the conclusion of therapy or who did not complete at least 18 months of therapy were excluded. This study was approved by the study location's institutional review board.

### *Objectives*

The primary objective was to compare the percent change from baseline in BMD between patients treated with Forteo® or PF708 teriparatide. Endpoints included BMD at baseline and after 18-24 months of therapy of the lumbar spine, left and right total hip, and left and right femoral neck.

The secondary objective was to compare safety and efficacy outcomes between Forteo® and PF708 teriparatide. Safety endpoints included reported side effects and change in serum calcium from baseline. Efficacy endpoints included incidence of vertebral, non-vertebral, or hip fractures while on treatment.

### *Data Collection and Analysis*

Patient profiles were identified using the health system's electronic medical record system (Epic®). Data was collected using RedCap® and de-identified prior to analysis. Data was analyzed using Microsoft Excel™. Only principal and sub-investigators had access to data files.

Data extraction included the following: baseline characteristics (age, sex, race, baseline body mass index (BMI), baseline BMD and T-scores (lumbar spine, left/right femoral neck, left/right total hip), baseline serum calcium (mg/dL), history of previous fracture, parent fractured hip, smoking status, use of glucocorticoids (daily dose of > 5 mg of prednisone or equivalent for > 3 months), history of rheumatoid arthritis, history of secondary osteoporosis (type 1 diabetes, osteogenesis imperfecta, untreated long-standing hyperthyroidism, hypogonadism or premature menopause, chronic malnutrition or malabsorption, chronic renal failure, chronic liver disease), alcohol consumption of ≥ 3 units per day, and type of DXA machine (Lunar or Hologic)), primary endpoints (BMD of lumbar spine, left/right femoral neck, and left/right total hip after 18-24 months of therapy), safety and efficacy secondary endpoints were collected during therapy (safety: reported side effects, first serum calcium during therapy (mg/dL); efficacy: first event of vertebral, non-vertebral, or hip fracture).

Baseline characteristics were analyzed using descriptive statistics. The primary endpoint was analyzed using Mann-Whitney test at significance level of 0.05 for non-parametric data. Bone mineral density data was analyzed for any femoral neck and any hip result to increase the size of the sample. For subjects with results for both the left and right side, the higher BMD value was used in the analysis. A sub-analysis was performed for the primary endpoint with removal of outliers. Secondary endpoints were analyzed using descriptive statistics.

**Results**

*Study Participants*

Between October 1, 2019 and October 31, 2022, 108 subjects met inclusion criteria. Of the 108 subjects, 27 were treated with PF708 teriparatide and 81 were treated with Forteo®. Baseline characteristics were similar in both subjects treated with PF708 teriparatide and subjects treated with Forteo®. Majority of the subjects were post-menopausal females (PF708 teriparatide 88.9%; Forteo® 80.2%) with a median age of 68 years (67 years; 68 years). Majority of subjects had a history of fracture prior to therapy initiation (PF708 teriparatide 81.5%; Forteo® 79.0%). Median baseline BMD in the lumbar spine (PF708 teriparatide 0.825 $g/cm^2$; Forteo® 0.867 $g/cm^2$), left femoral neck (0.689 $g/cm^2$; 0.662 $g/cm^2$), right femoral neck (0.671 $g/cm^2$; 0.682 $g/cm^2$), left hip (0.748 $g/cm^2$; 0.712 $g/cm^2$), and right hip BMD (0.698 $g/cm^2$; 0.760 $g/cm^2$) were similar between both groups. Additional baseline characteristics are provided in Table 1.

*Primary Objective*

Median percent change from baseline BMD of lumbar spine between PF708 teriparatide and Forteo®, respectively was +9.6% and +7.7% ($p = 0.56$), left femoral neck was +1.5% and +2.8% ($p = 0.39$), right femoral neck was +1.6% and +4.4% ($p = 0.14$), left total hip was +2.4% and +3.2% ($p = 0.24$), and right total hip was 0.0% and +1.4% ($p = 0.71$). Subjects with any femoral neck BMD, median change from baseline was +1.4% in the PF708 teriparatide group and +3.0% ($p = 0.03$) in the Forteo® group. Subjects with any hip BMD, median change from baseline was +1.6% and +2.6% between PF708 teriparatide and Forteo®. Median, interquartile range, and significance values in the outlier analysis remained consistent with the original sample (Table 4).

*Secondary Objective*

Adverse events occurred in 7.4% and 9.9% of subjects prescribed PF708 teriparatide and Forteo®, respectively. Adverse events reported were rash, fatigue, constipation, heart palpitations, joint pain, and nausea. Median change in serum calcium from baseline to first blood draw during therapy was +0.3 mg/dL in the PF708 teriparatide group and +0.2 mg/dL in the Forteo® group. Incidence of fracture during therapy was approximately 18.5% with PF708 teriparatide (18.5%) and 14.8% with Forteo® (Table 5).

## Discussion

Several teriparatide products have been approved for use in osteoporosis since the end of Forteo® market exclusivity. These products are designated as biosimilars or follow-on products depending on the approving country. PF708 teriparatide is manufactured by Alvogen and is designated as a drug, rather than a biologic, in the United States. Because of this, approval was allowed to be obtained without the need for safety and efficacy comparison with Forteo®, classifying it as a follow-on product. Comparison between Forteo® and PF708 teriparatide and changes in bone mineral density will provide real-world efficacy data to support the continued use of this product. Primary endpoint results of this study determined that differences in the change of bone mineral density from baseline were not statistically significant between PF708 teriparatide and Forteo®; a comparison that had not previously been investigated to our knowledge.

Results of this study are consistent with those published studies investigating other manufactured teriparatide products. Hadji P, et al. (2024) compared bone mineral density at the lumbar spine, femoral neck, and total hip at 12 and 24 months in patients treated with RBG-10 teriparatide manufactured by Gedeon Richter or Forteo®. Baseline characteristics of study participants were comparable to that of our study. No significant differences were observed between treatment groups in parameters of bone mineral density.[12]

In addition to bone mineral density changes, our study contributes safety data and incidence of fracture during therapy. PF708 teriparatide and Forteo® were well-tolerated in both groups with only 7.4% and 9.9% of subjects reporting adverse effects with PF708 teriparatide and Forteo®, respectively. Adverse effects in both groups were more likely to result in continuation or temporarily pausing therapy than discontinuation of therapy. Most common adverse events were rash, fatigue, constipation, heart palpitations, joint pain, and nausea. This data only includes patient-reported events documented in the electronic medical record.

Notable limitations of this study are its retrospective design and small sample size. The retrospective design limited data collection to information published in the electronic medical record. The small sample size was the result of restricting the study duration to include when PF708 teriparatide came to market and to be able to include at least two years' worth of data on each patient to reflect the recommended treatment duration for anabolic therapy. These limit both the strength and generalizability of the study findings. An additional limitation identified includes the variability in DXA scan results and DXA scan location. Not all patients included had data for every region identified in data extraction resulting in even smaller sample sizes when analyzing individual regions for percentage change in BMD. Also, not all patients had DXA scans completed at the study location or with the same DXA machine. There is potential for variability in results by location and machine.

Recommendations for future directions of research on this topic include larger scale, prospective trials with distinction between all available manufactured teriparatide products to determine therapeutic equivalence. Additionally, investigation of transitioning between Forteo® and teriparatide products during therapy to contribute data on interchangeability.

**Conclusion:**

This single-center retrospective chart review compared percent change from baseline in BMD after at least 18 months of therapy with Forteo® and PF708 teriparatide. The results of this study contribute real-world safety and efficacy data to support continued use of PF708 teriparatide for patients with osteoporosis at high-risk of fracture. Data supporting this specific teriparatide formulation had previously been lacking in the literature due to its approval pathway.

| Table 1: Baseline Demographic and Clinical Characteristics | | | |
|---|---|---|---|
| **Characteristic** | **PF708 Teriparatide (n = 27)** | **Forteo® (n = 81)** | **N = 108** |
| Age, median (IQR) | 67 (62.0 – 68.5) | 68 (62.0 – 72.0) | 68 (62.0 – 72.0) |
| Female, n (%) | 24 (88.9) | 71 (85.2) | 93 (86.1) |
| Postmenopausal female, n (%) | 24 (88.9) | 65 (80.2) | 89 (82.4) |
| Race, n (%) | | | |
| White/Caucasian | 25 (92.6) | 68 (84.0) | 93 (86.1) |
| Asian | 1 (3.7) | 7 (8.6) | 8 (7.4) |
| Black/African-American | --- | 3 (3.7) | 3 (2.8) |
| Not reported | 1 (3.7) | 3 (3.7) | 4 (3.7) |
| FRAX | | | |
| BMI, median (IQR) | 24 (20.2 – 26.6) | 22.7 (20.1 – 25.4) | 23 (20.1 – 25.6) |
| History of fracture, n (%) | 22 (81.5) | 64 (79.0) | 86 (79.6) |
| Parent fractured hip, n (%) | 7 (25.9) | 21 (25.9) | 28 (25.9) |
| Current smoker, n (%) | 1 (3.7) | 3 (3.7) | 4 (3.7) |
| Glucocorticoid use, n (%) | 3 (11.1) | 9 (11.1) | 12 (11.1) |
| History of rheumatoid arthritis, n (%) | 1 (3.7) | 1 (1.2) | 2 (1.9) |
| Secondary osteoporosis, n (%) | 5 (18.5) | 15 (18.5) | 20 (18.5) |
| Alcohol use, n (%) | 3 (11.1) | 11 (13.6) | 14 (13.0) |
| Type of DXA Machine, n (%) | | | |
| Lunar | 22 (82.1) | 62 (77.1) | 84 (78.4) |
| Hologic | 3 (11.1) | 12 (14.8) | 15 (13.9) |

| Table 2: Pre-Therapy BMD (g/cm$^2$), median (IQR) | | | | |
|---|---|---|---|---|
| | **PF708 Teriparatide (N = 27)** | | **Forteo® (N = 81)** | |
| Lumbar spine | n = 26 | 0.825 (0.753 – 0.883) | n = 73 | 0.867 (0.743 – 0.976) |
| Left femoral neck | n = 16 | 0.689 (0.555 – 0.819) | n = 46 | 0.662 (0.582 – 0.732) |
| Right femoral neck | n = 9 | 0.671 (0.624 – 0.691) | n = 39 | 0.682 (0.628 – 0.784) |
| Left hip | n = 18 | 0.748 (0.694 – 0.789) | n = 42 | 0.712 (0.653 – 0.788) |
| Right hip | n = 11 | 0.698 (0.674 – 0.767) | n = 35 | 0.760 (0.666 – 0.830) |
| Any femoral neck | n = 22 | 0.678 (0.613 – 0.805) | n = 66 | 0.680 (0.610 – 0.773) |
| Any hip | n = 24 | 0.737 (0.653 – 0.820) | n = 64 | 0.736 (0.653 – 0.820) |

| Table 3: Primary Endpoint– Percent Change in BMD, median (IQR) | | | | | |
|---|---|---|---|---|---|
| | PF708 Teriparatide (N = 27) | | Forteo® (N = 81) | | P-value |
| Lumbar spine | n = 26 | 9.6<br>(2.4 – 12.9) | n = 70 | 7.7<br>(4.1 – 11.9) | 0.56 |
| Left femoral neck | n = 16 | 1.5<br>(-1.9 – 5.8) | n = 41 | 2.8<br>(-1.3 – 6.9) | 0.39 |
| Right femoral neck | n = 8 | 1.6<br>(-0.01 – 3.9) | n = 36 | 4.4<br>(2.0 – 5.8) | 0.14 |
| Left hip | n = 18 | 2.4<br>(-1.1 – 3.9) | n = 39 | 3.2<br>(0.83 – 7.9) | 0.238 |
| Right hip | n = 9 | 0.0<br>(-1.3 – 1.6) | n = 34 | 1.4<br>(-2.1 – 4.0) | 0.71 |
| Any femoral neck<br>Teriparatide, n =22<br>Forteo, n=66 | n = 22 | 1.4<br>(-2.0 – 5.1) | n = 66 | 3.0<br>(0.93 – 6.7) | 0.03 |
| Any hip<br>Teriparatide, n = 24<br>Forteo, n = 64 | n = 24 | 1.6<br>(-1.6 – 4.4) | n = 64 | 2.6<br>(0.22 – 5.5) | 0.35 |

| Table 4: Primary Endpoint Outlier Analysis, median (IQR) | | | | | |
|---|---|---|---|---|---|
| | PF708 Teriparatide<br>(N = 27) | | Forteo®<br>(N = 81) | | P-value |
| Lumbar spine | n = 24 | 9.6<br>(2.7 – 12.9) | n = 70 | 7.7<br>(4.1 – 11.9) | 0.54 |
| Left femoral neck | n = 16 | 1.5<br>(-1.9 – 5.8) | n = 40 | 2.7<br>(-1.4 – 6.9) | 0.47 |
| Right femoral neck | n = 8 | 1.6<br>(0.0 – 3.9) | n = 34 | 4.4<br>(2.1 – 5.7) | 0.12 |
| Left total hip | n = 17 | 3.0<br>(0.6 – 4.1) | n = 38 | 3.3<br>(1.2 – 8.0) | 0.29 |
| Right total hip | n = 9 | 0.0<br>(-1.4 – 1.6) | n = 33 | 1.4<br>(-2.3 – 3.8) | 0.80 |
| Any femoral neck | n = 22 | 1.4<br>(-2.0 – 5.1) | n = 64 | 3.0<br>(1.1 – 6.2) | 0.03 |
| Any hip | n = 23 | 1.6<br>(-1.5 – 4.6) | n = 61 | 2.5<br>(0.3 – 5.4) | 0.57 |

| Table 5: Secondary Endpoints | | |
|---|---|---|
| **Endpoint** | **Teriparatide**<br>**n = 27** | **Forteo**<br>**n = 81** |
| Efficacy | | |
| Fracture during therapy, n (%) | 5 (18.5) | 12 (14.8) |
| Safety | | |
| Adverse effects, n (%) | 2 (7.4) | 8 (9.9) |
| Outcome of adverse effect, n (%) | | |
| Continuation of therapy | -- | 5 (6.2) |
| Discontinuation of therapy | 1 (3.7) | 3 (3.7) |
| Hold therapy | 1 (3.7) | -- |
| Change in serum calcium (mg/dL), median (IQR) | 0.3 (0.03 – 0.58) | 0.2 (-0.1 – 0.4) |